\documentclass[letter,10pt,twocolumn,titlepage]{IEEEtran}
\pdfoutput=1
\newcommand{\superscript}[1]{\ensuremath{^{\textrm{#1}}}}

\usepackage{graphicx}
\usepackage{subfigure}
\usepackage[sort]{cite}

\begin{document}
\title{Hard Data on Soft Errors: A Large-Scale Assessment of Real-World Error Rates in GPGPU}
\author{Imran S. Haque\superscript{1} \and Vijay S. Pande\superscript{1,2}\\
Stanford University, Departments of \superscript{1}Computer Science and \superscript{2}Chemistry}

\maketitle

\begin{abstract}
Graphics processing units (GPUs) are gaining widespread use in computational chemistry and other scientific simulation contexts because of their huge performance advantages relative to conventional CPUs. However, the reliability of GPUs in error-intolerant applications is largely unproven. In particular, a lack of error checking and correcting (ECC) capability in the memory subsystems of graphics cards has been cited as a hindrance to the acceptance of GPUs as high-performance coprocessors, but the impact of this design has not been previously quantified. 

In this article we present MemtestG80, our software for assessing memory error rates on NVIDIA G80 and GT200-architecture-based graphics cards. Furthermore, we present the results of a large-scale assessment of GPU error rate, conducted by running MemtestG80 on over 20,000 hosts on the Folding@home distributed computing network. Our control experiments on consumer-grade and dedicated-GPGPU hardware in a controlled environment found no errors. However, our survey over cards on Folding@home finds that, in their installed environments, two-thirds of tested GPUs exhibit a detectable, pattern-sensitive rate of memory soft errors. We demonstrate that these errors persist after controlling for overclocking and environmental proxies for temperature, but depend strongly on board architecture.

\end{abstract}

\section{Introduction}
Commodity programmable graphics hardware such as the AMD R600/R700 and NVIDIA G80/GT200 architectures have made available on the desktop TFLOP-scale floating-point performance formerly accessible only on dedicated supercomputers, with peak throughput improvements of over an order of magnitude relative to conventional CPUs. This extremely high performance makes GPUs attractive in computational chemistry, as many applications are bound by the limits of available computational power (e.g., the simulation time, simulated system size, and forcefield detail in molecular dynamics are all fundamentally constrained by available computation). Indeed, GPUs have been applied to several important problems in computational chemistry including molecular dynamics \cite{Friedrichs09,Stone07}, Poisson-Boltzmann electrostatics \cite{Narumi09}, DFT and MP2 quantum chemistry models \cite{Yasuda08,Ufimtsev08,Vogt08}, and molecular comparison \cite{Haque09}. Scientific problems outside chemistry, including biological sequence alignment \cite{Manavski08} and machine learning \cite{Catanzaro08} have also shown significant speedups from reimplementation for GPU execution. 

While GPGPU (general-purpose computation on GPUs) is attractive from the perspective of throughput, its origin in the relatively-error-tolerant area of consumer graphics has raised concerns about reliability. The reliability of GPUs in error-intolerant applications such as scientific simulations is largely unproven. Previous work \cite{Sheaffer06, Sheaffer07} has questioned the reliability of GPU logic and proposed methods for logic hardening. A particular point of concern, however, is the reliability of memory on GPUs. The lack of ECC (error-checking-and-correcting) functionality in GPGPU systems has been previously noted \cite{Sheaffer07}. 

While circumstantial evidence raises reliability questions, no prior work has actually quantified the reliability of GPGPU. Our contribution in this work is a quantification of the reliability of GPU memory subsystems in the context of GPGPU which has important implications for both developers and users of GPU-based scientific codes. We have carried out a large-scale assessment of GPU reliability by using a custom test code to measure error rates on more than 20,000 NVIDIA GPUs on the Folding@home distributed computing platform. In addition to the version of the tester deployed on Folding@home, we have released MemtestG80, a standalone version of our test code, under an open-source LGPL license at \texttt{https://simtk.org/home/memtest}.

Our experiment comprises over 200 terabyte-hours of memory testing distributed over more than 20,000 individual GPUs worldwide. We also present control experiments carried out on individual nodes and a GPU cluster. Our results demonstrate a detectable, pattern-sensitive rate of memory faults in the installed base of commercial GPU hardware. Specifically, even after controlling for overclocked cards and time of day of test execution (as a proxy for ambient temperature) we find that two-thirds of all tested GPUs exhibit sensitivity to memory faults in a pattern-dependent manner. This error rate depends strongly on board architecture, with devices based on the newer GT200 GPU exhibiting soft error rates nearly an order of magnitude lower than those based on the older G80/G92 design. 

Our paper is organized as follows. We first offer a primer on soft error rate generation and detection mechanisms, and explore prior work (Section \ref{sec:background}). We then present the design and validation  of MemtestG80, our software-based tester to detect soft errors in logic and memory on NVIDIA GPUs (Section \ref{sec:memtestG80}), and explain the methodology of our large-scale study of error rates using the Folding@home distributed computing network (Section \ref{sec:methodology}). In Section \ref{sec:results}, we present the results of our experiment. We subsequently present detailed analysis of the data in Section \ref{sec:analysis}, examining of possible error-inducing conditions. We conclude with a discussion of the implications of our findings on the field of reliable GPGPU and steps to be taken by both hardware and software developers.

\section{Background and Previous Work}
\label{sec:background}
Errors in hardware systems can be classified as ``soft'' or ``hard'': hard errors are triggered by physical hardware defects, whereas soft errors are random, transient faults not due to physical defects. The term ``soft error'' has traditionally referred to radiation-induced upsets in electronic circuits, in which high-energy particles from the environment (e.g., cosmic rays \cite{Ziegler96} or radiation from chip packaging materials \cite{Gordon08}) cause erroneous operation by creating localized charge disturbances \cite{Shivakumar02, Baumann02}. These soft errors are a significant problem in traditional supercomputing, as reflected in the error-checking-and-correcting design of the IBM BlueGene/L supercomputer \cite{Adiga02}. 

Mechanisms other than radiation can also induce transient errors. In memories, reads and writes can occasionally affect the state of adjacent cells; in both memories and logic, timing violations (e.g., from improper specification or overclocking) can cause transient errors \cite{Cockburn94}. Furthermore, at both the system and chip level, improper signal routing and power design can induce transient errors by various mechanisms such as voltage droop, crosstalk and timing jitter \cite{Sheaffer06, Metra00}. In this work, we use the term ``soft error'' to refer to the larger class of transient faults, rather than just radiation-induced errors.

While memory manufacturers typically do not disclose soft error rate (SER) information \cite{Cataldo01}, estimates have been drawn for RAM SER from a variety of sources \cite{Tezzaron04}, ranging from $5 \times 10^{-14}$ to $4 \times 10^{-7}$ errors per bit-hour. Data from IBM indicate that in natural environments, even with hundreds of devices under test, more than 1,000 testing hours may be required to accumulate statistically meaningful test results \cite{Gordon08}. Additionally, possible environmental (e.g., thermal or radiation) effects on SER dictate that a variety of conditions be tested. For example, cosmic ray flux varies by a factor of two depending on latitude \cite{Ziegler96}, and approximately 13x between sea level and 10,200 ft in altitude \cite{Gordon08}. Thus, a very-large-scale approach is required in order to efficiently accumulate statistically-significant data regarding SER.

Because of the significance of SER to the semiconductor and computer industries, much past work has been done on techniques for measuring memory and logic SER. Cockburn \cite{Cockburn94} presents an introduction to hardware testing methods and fault models used by semiconductor device manufacturers to test memories, and also presents the use of random test patterns as a simple method that achieves good test coverage across a variety of failure modes. Suk and Reddy \cite{Suk80} establish the importance of using different test patterns to detect pattern-sensitive device faults; however, their presented patterns depend on intimate knowledge of the underlying memory layout. 

These hardware-based testing methods are useful for verifying sensitivity of hardware to design or manufacturing flaws. However, the low radiation flux in most (habitable) environments makes this sort of testing impractical to detect radiation-sensitivity faults in a high-throughput fashion. Consequently, semiconductor manufacturers will sometimes use high-radiation test environments to further characterize devices. Past work includes the use of radioactive isotopes \cite{Blandford85} and particle accelerators \cite{Violante07} to bombard chips with high levels of radiation. An alternative approach is the use of high-altitude testing \cite{Gordon08} (cosmic ray neutron flux rises exponentially with altitude \cite{Ziegler96}).

Higher-level software and modeling-based techniques have also been proposed for soft error detection and correction. Software hardening techniques can be used to detect logic and memory soft errors \cite{Nicolescu03, Nicolescu01}. Walcott et al.\ describe a method for predicting the vulnerability of logic architectures to soft error and the ``architectural vulnerability'' \cite{Mukherjee05} of different codes running on the same architecure \cite{Walcott07} and validate their approach through software simulation. Sheaffer et al.\ also take a simulation-based approach and specifically characterize the architectural vulnerability of graphics algorithms on GPUs \cite{Sheaffer06}. As their later work points out, however, this characterization is inappropriate for GPGPU, which requires tighter error guarantees than does conventional graphics \cite{Sheaffer07}.

While prior work has laid out testing methods to detect soft errors, no previous studies have applied these techniques to a large installed base to broadly assess the impact of soft errors on the emerging GPGPU platform. Our contributions in this work are twofold: first, a test code using proven memory and logic testing methods to detect soft errors; second, a large body of data using this tester to assess SER on tens of thousands of installed GPUs worldwide.

\section{Design and Validation of MemtestG80}
\label{sec:memtestG80}
In this section we describe the design and validation of Mem\-testG80, our CUDA-based \cite{Nickolls08} code to test for memory errors on NVIDIA GPUs based on the G80/Tesla architecture \cite{Lindholm08}. MemtestG80 is a CUDA implementation of most of the memory tests in Memtest86 \cite{Memtest86}, a widely-used open-source memory tester for x86-based machines that implements many popular memory-test patterns, including randomized tests and tests for pattern-sensitive errors. Section \ref{sec:testMI} briefly lists the tests implemented in MemtestG80. For conciseness, we refer the reader to the Memtest86 documentation \cite{Memtest86} for descriptions of most of the test patterns, and here highlight instead the design decisions made in a parallel GPU implementation of the code. We also detail the implementation of a logic test of our own devising that is unique to MemtestG80. 

In the text that follows, one ``iteration'' of MemtestG80 refers to the execution of each implemented test once. For tests that take place in multiple rounds, every round is executed (with one exception, explained in Section \ref{sec:procedure}). Such an iteration over 64 MiB of memory typically takes between 1 to 5 seconds to complete, depending on GPU speed; runtime scales linearly with the amount of memory tested.

\subsection{Offload and Parallelization Scheme}

Several design parameters affect the sensitivity and speed of a software-based GPU memory tester. Specifically, the three components of the memory tester --- pattern generation, memory access (writing and reading patterns to/from memory), and pattern verification --- can be performed either by the CPU or the GPU itself; and if performed on the GPU, can be performed either serially or in parallel across the multiple GPU cores. The decisions made in MemtestG80 are informed both by the assumptions we make about the relative error rates of various system components and by responsiveness requirements dictated by operation on donated distributed-computing resources. 

To improve the speed and responsiveness of the memory tester, all pattern generation, memory access, and verification is done in parallel on the GPU. We assume that the memory error rate is sufficiently low that the on-GPU code (which occupies a much smaller footprint than the tested region) will not be corrupted during the test execution. Performing verification on the GPU leaves the tester vulnerable to GPU logic errors. We therefore implicitly assume that the GPU logic error rate is lower than the GPU memory error rate; however, we verify this assumption by also running a custom logic test that should report errors on architectural paths similar to those used in other parts of the tester. 

\subsection{Logic Testing}

Because results from the GPU can be passed back to the host CPU only by a copy from the GPU main memory, detection of GPU logic errors under the assumption that memory errors are more frequent than logic errors is nontrivial --- an error in a computed result may be caused by an error in logic or memory. To overcome this problem, a test can be designed which produces the same expected result after varying amounts of logic operations. The same test can be run twice with (for example) four times the number of logic operations in the second execution. Since both tests write the same data to memory, the expected rate of errors due to memory faults will be equal between the two executions; since the latter test runs more logic operations, errors from logic faults should scale with the number of operations.

The design of our logic test, unique to MemtestG80, is based on the preceding principle. For the core calculation, we use a linear congruential random number generator (LCG) with a short period $k$ starting from zero. Such a generator, when started from zero, will return to zero after a fixed number of iterations $k$. Because the generator only reaches $k$ states, of $2^{32}$ possible (in the 32-bit generator), assuming a uniform probability of error over bits, most logic errors will cause the generator to transition to a state outside the normal operation cycle. Such a state is unlikely to return to zero in the correct number of steps, and therefore whether the generator returns to zero is a good indication of whether a logic error occurred. Our logic test starts the generator from zero and runs it for $k$ or $4k$ cycles, each time writing the results out to memory, reading it back, and verifying that it contains only zeros. Any nonzero values indicate either the presence of a logic or a memory error. Scaling of the number of nonzero values with the number of LCG iterations indicates logic, rather than memory, errors. The use of constant zero as the test pattern further increases the sensitivity of the test to logic rather than memory errors; as we show in Section \ref{sec:validation}, the constant zero pattern is insensitive to faulty memory.

\subsection{Validation}
\label{sec:validation}
To validate MemtestG80, we carried out both negative and positive control experiments. Since the purpose of this study is to investigate the latent error rate in GPU hardware, a true negative control (one in which we are guaranteed no errors) is not possible. To minimize the possibility of errors from overheating or power disturbances, the controls were run on machines in controlled-temperature environments with known-good power sources. Shielding against ambient radiation such as cosmic rays requires meters of concrete or rock \cite{Tezzaron04} and was therefore deemed impractical.

We ran negative controls on two types of hardware. The first was a GeForce 8800 GTX, a high-end consumer-grade graphics board operating at NVIDIA reference clock frequencies. Secondly, we ran tests on a cluster of 8 NVIDIA Tesla C870 boards. The Tesla C870 is a board designed specifically for GPGPU, is based on the same GPU architecture as the 8800GTX, and should reflect any engineering changes made by NVIDIA for ``pro'' or GPGPU-grade cards. Over 93,000 iterations of MemtestG80 were run over 700MiB of GPU memory on the 8800GTX. An aggregate 1.48 million iterations were run over 320 MiB on each Tesla board. No errors were ever detected on the negative control 8800GTX or Tesla boards, demonstrating that errors detected by MemtestG80 are unlikely to be spurious or due to errors in the code.

\begin{figure*}[ht]
\centering
\includegraphics[height=0.33\textheight]{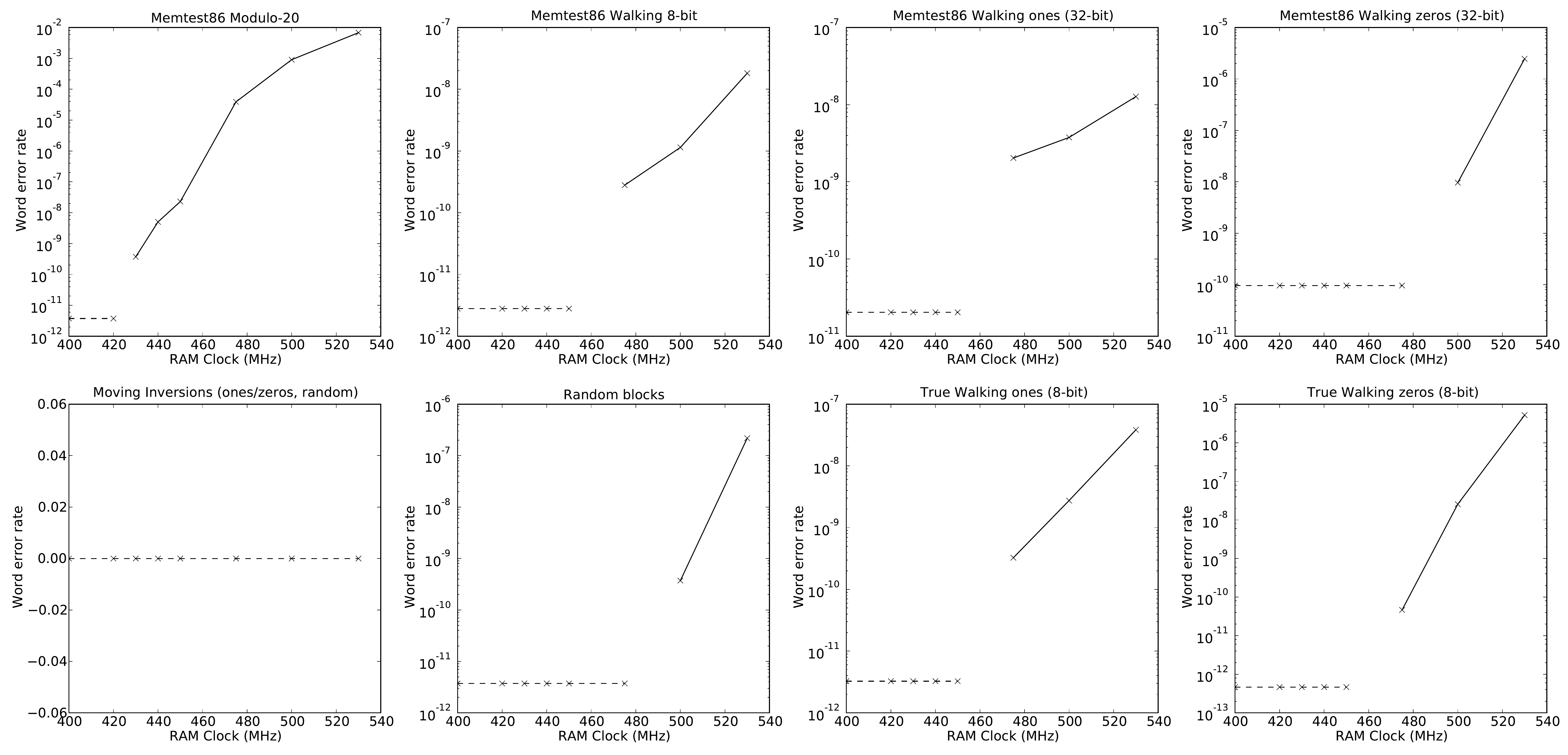}
\caption{Overclocking positive control experiment: word error rate (ratio of incorrect to total words tested) versus clock rate for each memory test type. Both moving inversions tests displayed together as  neither ever reported an error. Dashed lines represent zero errors found at the tested frequency and are arbitrarily set two log units below lowest number of errors. }
\label{fig:testSensitivity}
\end{figure*}

To ensure that MemtestG80 detects errors under conditions known to generate memory errors, we also carried out a positive control experiment. Since overclocking is known to generate memory errors (due to violations of timing constraints on both memory and memory controller circuitry), we used overclocking as our positive control. MemtestG80 was run on a GeForce 9500GT video card (default memory clock rate = 400 MHz) with its memory clock rate set to 400, 420, 430, 440, 450, 475, 500, and 530 MHz. Each test comprised 20 iterations of MemtestG80 (10 at 530MHz because of prompt instability). Between each clock frequency the board was reset to a memory clock of 400MHz and allowed to cool to a constant temperature to avoid thermal effects from affecting test results. We finally ran an additional 20 iterations at 400 MHz to verify that the results were unaffected by the intermediate overclocking. The results of the positive control experiment are presented in Figure \ref{fig:testSensitivity}.

Two results deserve special attention. First, both variants of the moving-inversions test (which write the same constant --- zero, all ones, or a random number --- to all words in tested memory, and read it back) are completely insensitive to overclocking-induced errors. This inspires our choice of the all-zero pattern as the logic test pattern, as it seems to be insensitive to memory errors. 

Second, the modulo-20 test is far more sensitive to over\-clock\-ing-induced errors than are the other tests, demonstrated by the fact that it started detecting errors at clock rates lower than those required to trigger errors in other tests. The modulo-20 test proceeds in 20 rounds. In round $i$, a 32-bit pattern is written to each memory location whose offset from the start of tested memory is equal to $i$ modulo 20; the bitwise complement of the pattern is then written twice to every other memory location. In the verification kernel, the offsets equal to $i$ modulo 20 are read back and it is verified that they contain the original pattern. This procedure is repeated for $i \in \{0,1,2,\cdots,19\}$. 

The sensitivity of this test likely stems from the fact that the test's stride of 20 is unlikely to align with any architectural stride in the memory chips (e.g., a row length or chip interleave factor), so test locations are more likely to be physically scattered and to be influenced by neighboring cells. The sensitivity of the modulo-20 test persists in real-world situations and is key to our sampling of error rates. The pattern sensitivity of the memory subsystem is in itself an interesting result, as it demonstrates that the likelihood of encountering a memory error is highly dependent on the access and data patterns.

\section{Experimental Methodology}
\label{sec:methodology}
\subsection{Testing Procedure}
\label{sec:procedure}
To carry out the large-scale assessment presented in Section \ref{sec:results}, MemtestG80 was deployed on the Folding@home (FAH) distributed computing network \cite{ShirtsPande00,Beberg09}. For each execution of MemtestG80, we collected device information (card name, memory size, and shader-domain clock speed). Because of the widely varying capabilities of GPUs on Folding@home, the size of the tested memory region varied between 32 and 128 MiB; sizes larger than 128MiB were disallowed because of the negative impact on the responsiveness of donors' computers. Because of the high memory bandwidths required on GPUs, it is likely that memory blocks are interleaved across physical memory chips to speed total throughput. We believe that the tested memory region sizes are sufficiently large that they are likely to be spread across all or a substantial fraction of physical chips on the tested devices. The LCG period used in the logic test was 256, 512, or 1024; like the size of the test region, the LCG period varied based on the capability of donor graphics cards. Finally, on Folding@home, only 2 rounds of the modulo-20 test were run per test iteration because of responsiveness concerns. Later rounds were performed in following iterations.

On these test regions we ran a variable number of test iterations, collecting individual results for every test iteration. Rather than measuring the exact bit-error-rate (which may be unreliable due to GPU logic errors), we check only whether \emph{any} errors were detected during a test iteration. This ensures the robustness of the test. It is possible that a test which should have returned errors may have its ``error flag'' bits toggled off by a subsequent GPU memory or logic error and will therefore not be detected, creating a false negative. If a test successfully executes, it is possible that a GPU error will toggle an error flag bit on --- but this is in itself a GPU error. Thus, this binary decision approach removes the problem of false positive results. In the worst case, our results will underestimate the error rate; they will never overestimate it.

\subsection{Test Device Statistics}

The test was run at least once on over 20,000 distinct GPUs, with an aggregate total of over 4.6 billion test iterations executed. 96.6\% of our data were collected with a test region size of 64 MiB and an LCG period of 512; 3.2\% were collected with 32 MiB regions and period-256 LCGs, and the remainder with 128 MiB test regions and period-1024 LCGs. The tested boards are distributed worldwide with excellent coverage of North America and Europe; distributions elsewhere cluster sharply near large population centers (however, besides South Africa, Africa is poorly sampled).

Using NVIDIA specifications and the shader clock speeds reported by each board, we are largely able to classify boards as overclocked or running at ``stock'' frequencies. In a few cases, identically-named boards have multiple possible stock clock rates (e.g., models that fit into different thermal envelopes); in these cases it is not possible to determine whether or not a board is overclocked. Although memory clock rates will have a larger impact on memory error rate than logic (shader) clock frequencies, memory overclocking can be expected to covary with shader overclocking. The typical reason for Folding@home users to overclock their boards is to increase their throughput on Folding@home molecular dynamics (MD) work units. Empirical testing has shown that shader clock frequencies have more of an impact on MD runtimes than memory clocks, so it is unlikely that a board on Folding@home would have overclocked memory but not overclocked shaders. Some cards are shipped by board vendors in an overclocked state; here too, it would be rare to see overclocked memory without overclocked shaders, as shader clock frequencies can have a major impact on graphics performance. 

\begin{figure}[t]
\centering
\includegraphics[trim=1cm 5mm 1cm 8mm,clip,width=\columnwidth]{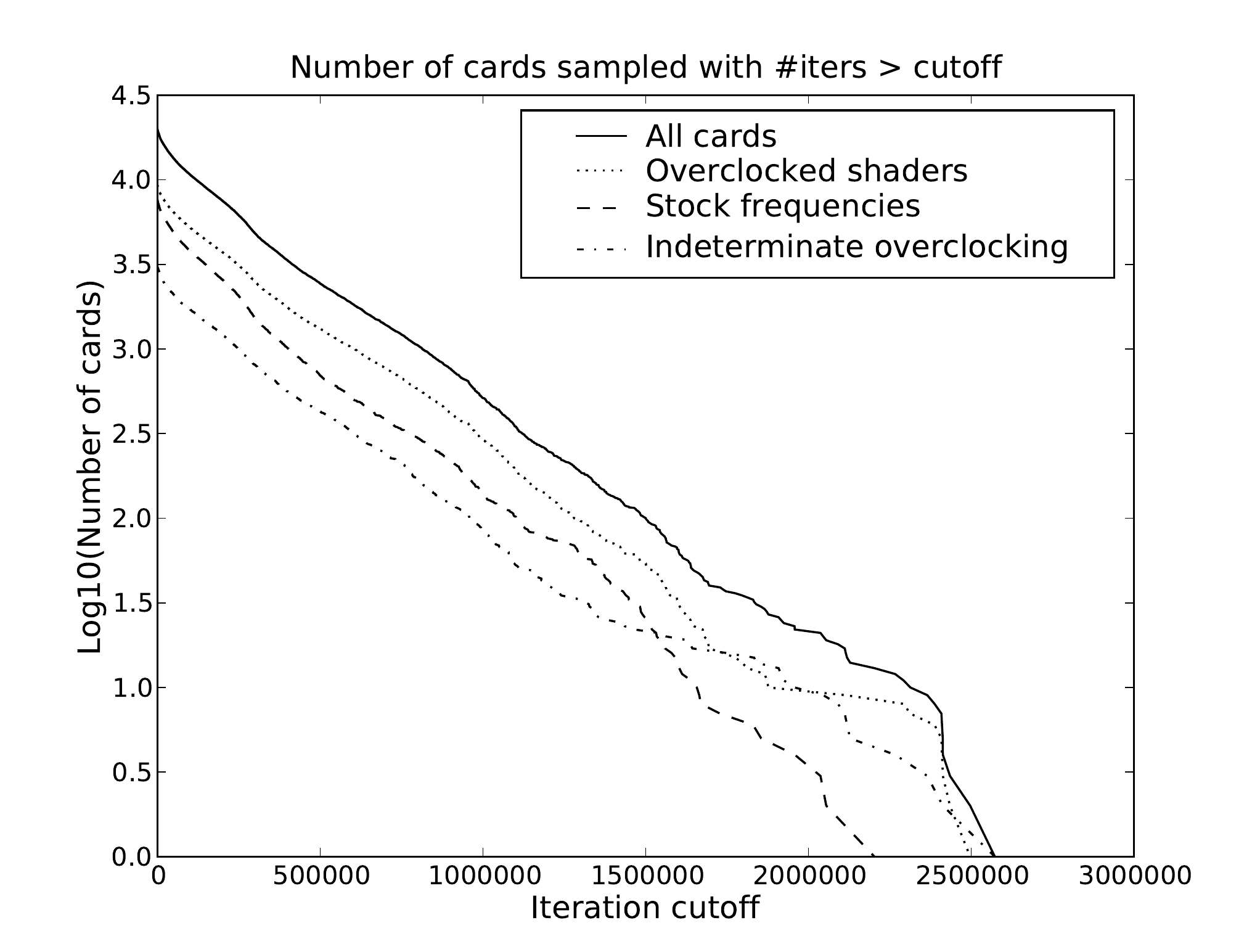}
\caption{Number of cards that completed at least a given number of test iterations}
\label{fig:NcardsVcutoff}
\end{figure}

Figure \ref{fig:NcardsVcutoff} displays the number of cards that completed a given number of MemtestG80 iterations during the course of this experiment. It further breaks the data into cards running at or below stock frequencies (grouped together as ``stock''), overclocked cards, and cards whose overclocking state is indeterminate. The results show that a slight majority of cards on Folding@home are overclocked; for most iteration count cutoffs, the number of overclocked and stock cards is comparable. 

Finally, we achieve good coverage of GPUs across the NVIDIA product line. Table \ref{tab:cards300k} shows the counts of cards in particular NVIDIA product families that completed at least 300,000 iterations during the course of the experiment. Although the dataset is strongly biased towards NVIDIA consumer graphics cards (GeForce) due to the consumer-oriented nature of Folding@home, we do sample a few professional (Quadro) and GPGPU-dedicated (Tesla) boards.

\begin{table}[t]
\centering
\begin{tabular}{| c | c |}
\hline
\textbf{Card Family} & \textbf{\# cards $\geq$ 300,000 iter.}  \\ \hline \hline
\textit{Consumer graphics cards} & \textit{4754 total} \\ \hline
GeForce 8800 & 1779  \\ \hline
GeForce 9800/GTS & 1375  \\ \hline
GeForce GTX  & 1137  \\ \hline
GeForce 9600  & 368  \\ \hline
GeForce 8600  & 65  \\ \hline
GeForce 9500  & 19  \\ \hline
Mobile GeForce & 11 \\ \hline \hline
\textit{Professional graphics cards} & \textit{36 total} \\ \hline
Quadro FX & 29  \\ \hline
Quadroplex 2200 & 6  \\ \hline
Quadro NVS & 1  \\ \hline \hline
\textit{Dedicated GPGPU cards} & \textit{22 total} \\ \hline
Tesla T10 & 20  \\ \hline
Tesla C1060 & 2  \\ \hline
\end{tabular}
\caption{Distribution of cards tested on FAH that ran at least 300,000 test iterations}
\label{tab:cards300k}
\end{table}

\section{Results}
\label{sec:results}
We classified each test iteration returned as having failed or not using the method described in Section \ref{sec:procedure}. We then inferred an empirical probability of failure (in a given test iteration) for each card tested as the ratio of failed tests to total tests, thereby estimating an empirical probability distribution that any card might have a given probability of failure. To add statistical validity, we applied various cutoffs for the minimum number of test iterations a card must have completed to be used in constructing this empirical card-reliability probability distribution.  Our underlying statistical model is that each card (in combination with its environment) has its own probability of failure $\mathit{P\left( fail \right)}$, and that each card is drawn independently from some underlying distribution $\mathit{P\left( P \left( fail \right) \right)}$. In all following plots and analyses, ``$\mathit{P\left( fail \right)}$'' refers to any given card's probability of failing a single MemtestG80 iteration.

\begin{figure*}[t]
\centering
\subfigure[Empirical CDFs of card failure probability at several test-iteration-count thresholds] {
\includegraphics[width=0.95\columnwidth]{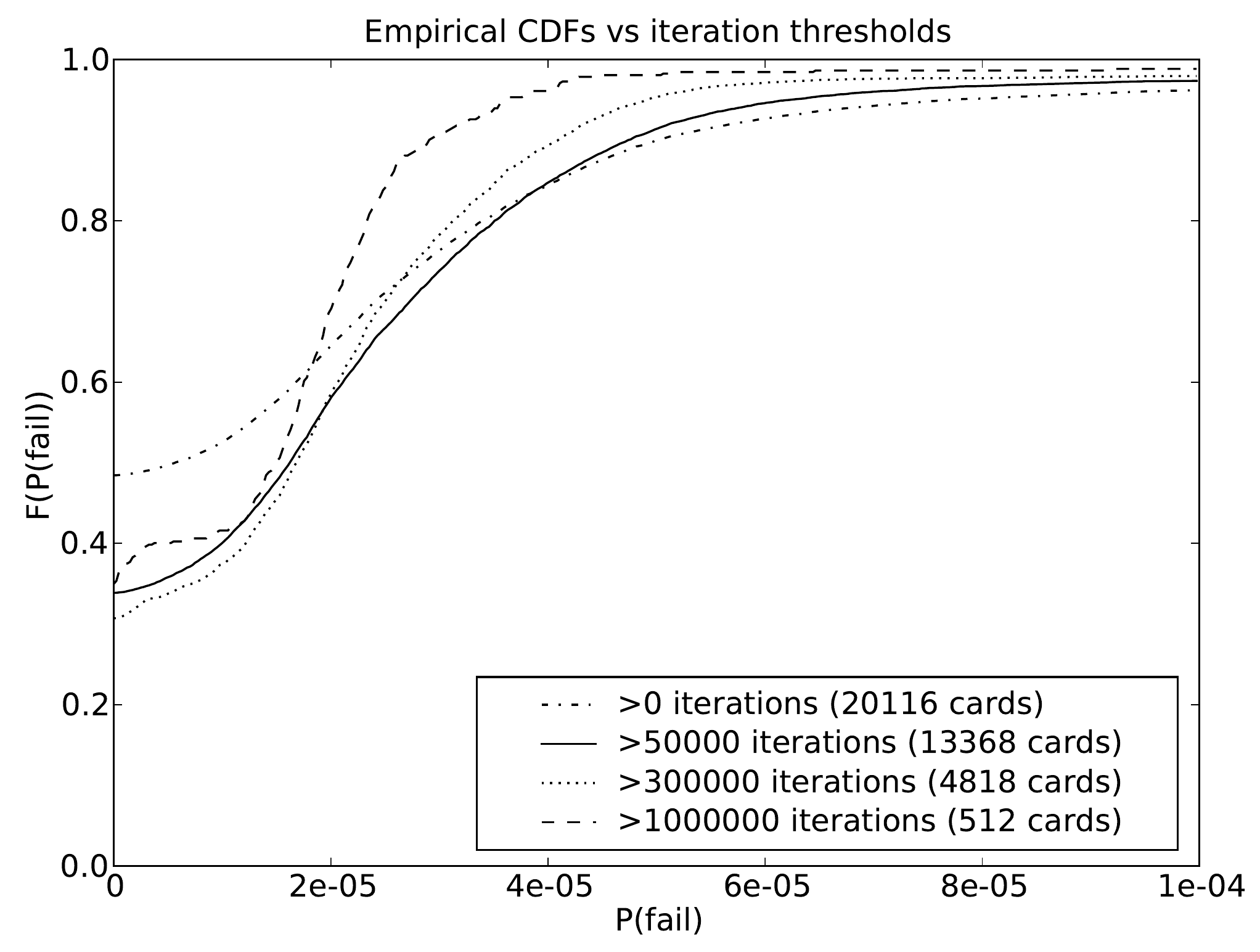}
\label{fig:PfailCDF_few}
}
\subfigure[Empirical PMF and CDF of card failure probabilities for cards running at least 300,000 MemtestG80 iterations] {
\includegraphics[width=0.95\columnwidth]{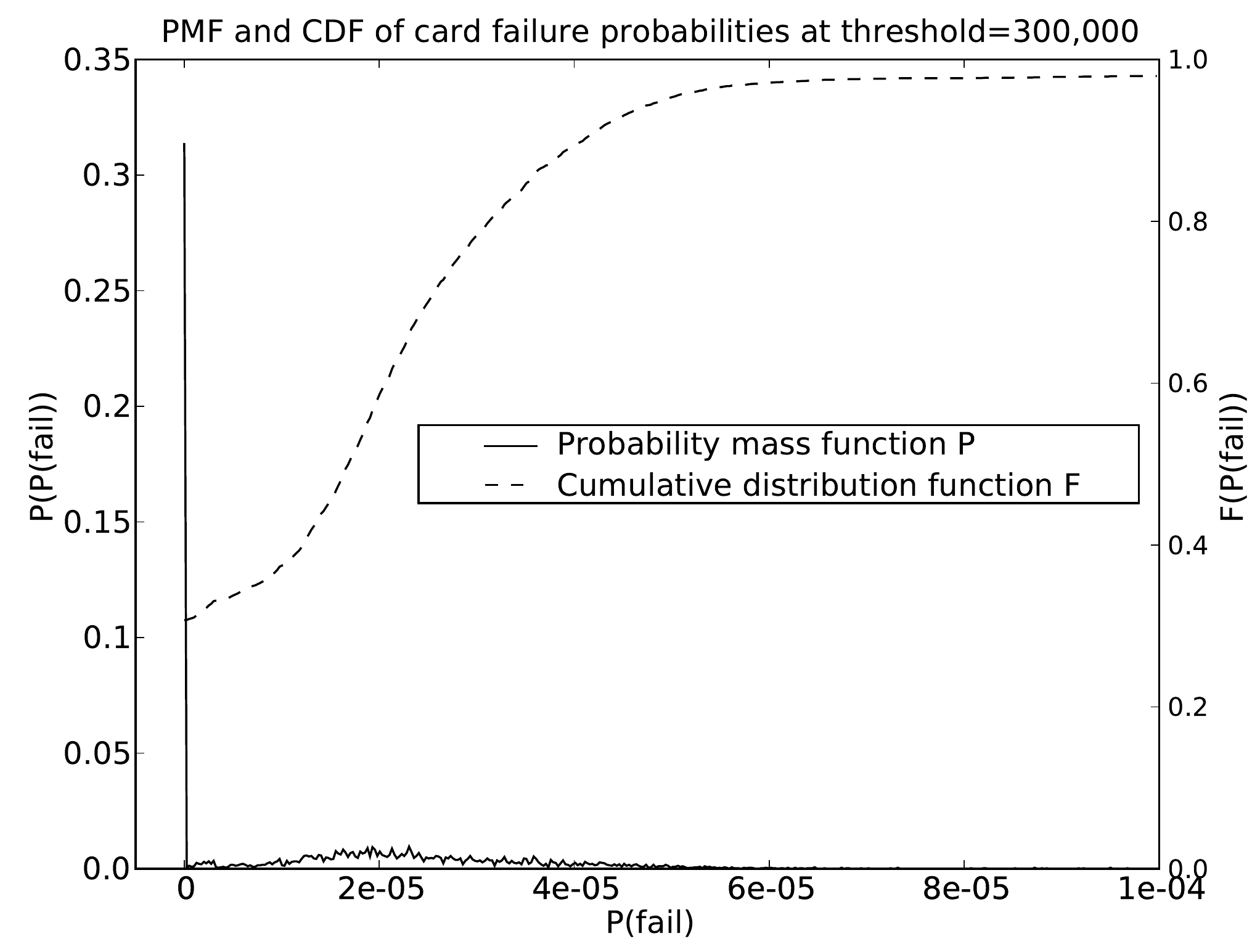}
\label{fig:Pfail_PMFCDF_300k}
}
\caption{Empirical probability mass functions (PMFs) and cumulative distribution functions (CDFs) of card failure probabilities}
\end{figure*}

Figure \ref{fig:PfailCDF_few} displays the cumulative distribution functions (integrated probability distributions) derived from this data for 4 values of the iteration threshold. Each trace represents the distribution calculated using a different cutoff for the number of iterations required to have been completed to consider a card for inclusion. A larger number of completed iterations for a card increases the statistical certainty that its probability of failure lies in the given bin of the estimated probability distribution. We present cutoffs only up to 1 million test iterations because the number of cards sampled falls off rapidly past this limit; estimates made using only cards past a cutoff beyond this are not statistically useful.

The most apparent trend in the data is the strongly bimodal distribution. All the CDFs start with a nonzero value at $\mathit{P\left( fail \right)} = 0$, representing the fraction of cards at each threshold which never failed a test. All CDFs further show a second population with a mean $\mathit{P\left( fail \right)}$ around $2 \times 10^{-5}$, which represents nearly all the remaining cards. Finally, there is a very small population of cards with failure rates higher than $1 \times 10^{-4}$, likely representing faulty hardware. This bimodal trend is statistically relevant, as it continues to appear in the data even at the largest cutoff. In particular, the distributions at thresholds of 50,000, 300,000, and 1,000,000 iterations all have similar fractions of cards with zero errors, indicating that this particular population is stable (i.e., measuring with more iterations does not find errors from the zero-error population).

As these trends are stable with respect to iteration cutoff, it is instructive to examine the distribution of failure probabilities at a single, representative cutoff that maintains statistical validity. Figure \ref{fig:Pfail_PMFCDF_300k} illustrates both the probability mass function and the cumulative distribution function over failure probabilities at an iteration cutoff of $\geq$ 300,000 iterations. At this threshold, approximately one-third of cards tested never exhibited a memory error. Nearly all of the remainder had failure probabilities between 0 and $10^{-4}$; only about 2\% had failure probabilities higher than this.

\section{Analysis}
\label{sec:analysis}
In this section we explore various hypotheses explaining features of the returned results, breaking the results down by test and by properties of the tested cards. The main statistical methods we apply are an examination of the mutual information between two probability distributions, and the information gain criterion for data partitioning attributes. Colloquially speaking, the mutual information between two distributions is a nonlinear measure of correlation between the two, and the information gain in an attribute measures the amount of variability in an underlying distribution that is explained by the attribute. These techniques are well-known in the statistical literature \cite{CoverThomas, WittenFrank}.

\subsection{Hypothesis testing by information gain}
\label{sec:hypotest}

To test our hypotheses we apply the information-theoretic measure known as \emph{information gain}, which is broadly used in data mining as a heuristic criterion for building decision tree models of data \cite{WittenFrank}. The hypothesis testing problem is formulated as follows: given a labeled dataset $D$, we partition $D$ according to an indicator variable $V$ into multiple subsets $D_1, D_2, \cdots, D_{|V|}$. We would like to know how good $V$ is at explaining the variability in $D$. 

We measure the ``variability'' of D and each of its subsets by their respective Shannon entropies, $H(D)$, defined as
\begin{displaymath}
\nonumber H(D) = -\sum_{x \in D} p(x) \log _2\left( p(x) \right)
\end{displaymath}
The information gain on $D$ from $V$, $I(D;V)$ (also known as the mutual information between $D$ and $V$), is  defined as:
\begin{eqnarray}
\nonumber I(D;V) &=& H(D) - H(D|V) \\
\nonumber I(D;V) &=& H(D) - \sum _{v \in V} H(D_v)P(V = v)
\end{eqnarray}

If $I(D;V)$ is large compared to $H(D)$, then $V$ explains a significant portion of the distribution of $D$. 

To estimate probability distributions $D$ in our hypothesis testing, we histogrammed failure probabilities on a per-card basis as was done for each distribution in Figure \ref{fig:PfailCDF_few}, but across the entire range of probabilities from 0 to 1. Although this resolution is too high for the low number of counts at higher probabilities, most of these bins will be zero-valued and will not affect the entropy calculations.

\subsection{Bimodality of P(fail)}
The bimodal distribution of card failure probabilities illustrated in Figure \ref{fig:Pfail_PMFCDF_300k} raises an obvious question: is the existence of cards with nonzero failure probability easily explained through a simple structural or environmental variable, or is it inherent to population of boards? To answer this we tested a set of obvious hypotheses on our data set: that failures are due to overclocking, that they are caused by thermal problems, or that they reflect architectural variability in the hardware.

As a reference, a perfect indicator variable on this dataset (separating the data into cards which never failed and those which did) has an information gain $I(D;V)$ of 0.8896 bits.

\subsubsection{Shader Overclocking}

To test whether shader-overclocked cards are responsible for the second mode in the distribution, around $\mathit{P\left( fail \right)} = 2 \times 10^{-5}$, we let $D$ be the set of all cards with a known overclocking status, and partitioned it into known stock and known overclocked cards. The entropy of $D$ was calculated to be 6.184 bits, and the mutual information between $D$ and the stock/overclocked indicator variable was $7.5 \times 10^{-2}$ bits. $I(D;V)$ for this split is much lower than that for a perfect indicator, indicating that overclocking status explains very little of the information in the distribution. Thus, it is unlikely that overclocking is the cause of bimodality.

\subsubsection{Time of day}

High temperatures are a known cause of transient errors in electronics. Although the APIs used by MemtestG80 do not allow us to monitor board temperatures, the status of Folding@home as a donor project suggests that most tested boards will not be in closely-temperature-controlled environments such as machine rooms. We combine the estimated test runtime, time results were received, and IP geolocation data \cite{GeoLite} to estimate the local time of day during the. Assuming that the ambient temperature will fluctuate with local time of day, we test the hypothesis that time of day controls the shape of our error distribution. 

We define ``day'' to run from 6am to 6pm, and ``night'' as 6pm to 6am, local to where the test was run. We let $D$ be the set of all tests which both started and ended during the day or during the night, and split it into tests that ran exclusively during the day and tests which ran exclusively at night. This hypothesis has an information gain of 0.0413 bits, indicating that it is exceedingly poor at explaining the shape of the error distribution; local time of day is therefore not an adequate explanation for error rates.

\subsubsection{Board architecture}

The GT200 series GPU from NVIDIA has a more advanced memory controller than the original G80/G92 GPU, supporting (perhaps among other redesigned features) additional memory coalescing modes. To test whether this change in GPU architecture was the cause of bimodality, we let $D$ be all GeForce graphics cards, and partitioned the set into GT200-based and non-GT200-based boards. $I(D;V)$ for this indicator was 0.453 bits, which is a large fraction of the information gained from a perfect indicator. Thus, it is likely that this division significantly explains the error distribution we observe. 

\begin{figure}
\centering
\includegraphics[width=\columnwidth]{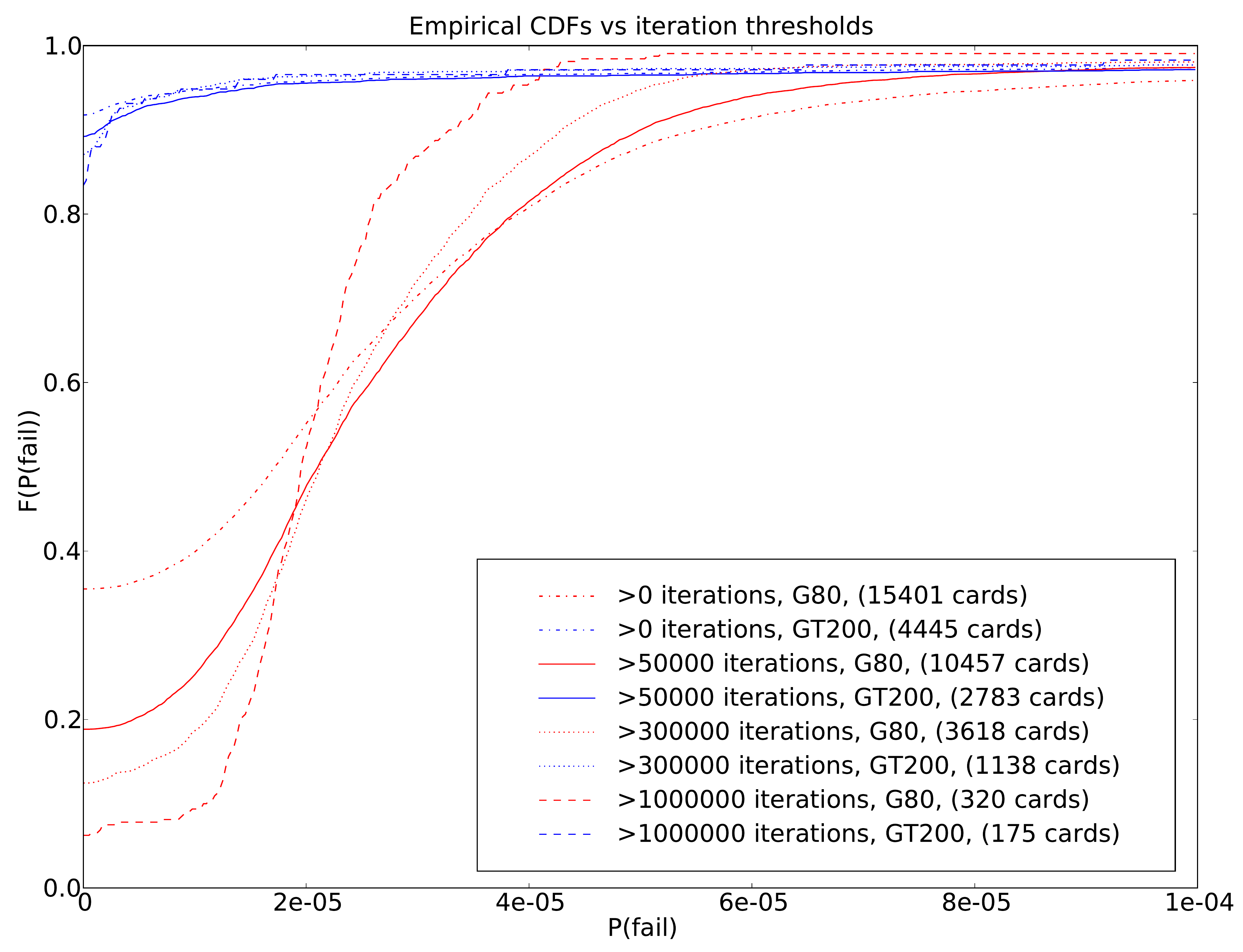}
\caption{Empirical CDFs of card failure probability at several test-iteration-count thresholds, by architecture (G80/G92 in red, GT200 in blue)}
\label{fig:PfailCDF_few_byarch}
\end{figure}

Figure \ref{fig:PfailCDF_few_byarch} shows that GT200-based boards (comprising about one-fifth to one-fourth of our dataset, depending on iteration threshold) were far less likely to fail MemtestG80 iterations --- regardless of iteration threshold, approximately 90\% of GT200-based cards never reported any errors. The population of GT200-based boards producing errors clusters at a failure probability of $2.2 \times 10^{-6}$, an order of magnitude lower than the mode failure probability for the overall dataset.

\subsubsection{Bimodality conclusions}

Our data suggest that the bimodal structure of the failure probability distributions is caused by differing architectures in the boards tested. Specifically, the newer GT200 architecture has an apparent soft error rate nearly tenfold lower than that of G80. The most obvious user-visible enhancement on the GT200 memory controller relative to that on G80 is improved support for coalescing memory operations, or combining multiple memory reads or writes into single transactions. However, this support is insufficient to explain the change in error rates. 

Using the published guidelines for memory coalescing on either architecture \cite{CudaGuide2.2}, we simulated the memory access patterns for G80 and GT200 on the modulo-20 test (the most sensitive test). As performed in MemtestG80 on Folding@home, the modulo-20 test executes the same number of transactions on both architectures. However, GT200 is able to shrink transactions to be smaller (in terms of bytes) than those performed on G80. As a consequence, G80 generates 16.7\% more memory traffic (in terms of bytes, not transactions) than GT200 on the test. By itself this does not appear to explain a 10-fold reduction in error probability by GT200. 

The large sample sizes for boards on both architectures make it unlikely that there is a consistent environmental difference between installations of either board type. While it is possible that the error rate is an age-induced effect (G80 is an older architecture than GT200, and it is possible that G80 boards in our sample are physically older than GT200 boards), our data seem to indicate that GT200 is in fact more resistant to soft error generation than is G80.

\subsection{Impact of errors on molecular dynamics}

To assess whether memory errors have an impact on scientific computing, we looked for mutual information between the probability that a given card generates memory errors on its Folding@home work units and the probability that the same card triggers an ``early unit end'' (EUE), or simulation failure, on its Folding@home work units. Counting only work units in which at least one MemtestG80 iteration was executed, the mutual information between MemtestG80 errors and Folding@home EUEs was 0.131 bits, compared to overall distribution entropies of 1.965 and 1.018 bits respectively for memory error and EUE distributions. This indicates that MemtestG80 errors likely do not correlate well with EUEs. However, we believe that this measure underreports the true impact on the simulations. EUEs have a variety of causes (including improper simulation setup) which may be unrelated to errors on the board; furthermore, because of the design of the Folding@home client used, certain simulation errors were not reported to the servers and could not be logged. Hence, our results are inconclusive as to the true impact of observed errors on scientific simulations.

\subsection{Failure modes of tests}
\label{sec:testMI}

By examining the mutual information between the results of each individual test comprising a MemtestG80 iteration, it is possible to better understand the mechanisms triggering failures under various conditions. For each test, we construct a list in which each element corresponds to a single execution of MemtestG80, and the value of each element is the number of failures on that test for that execution. Corresponding elements in each vector map to the same MemtestG80 execution. Each list of failure counts was then histogrammed into 10 bins and normalized to build an empirical probability mass function for the number of failures in that test on a given execution of MemtestG80. Using these probability distributions for tests $X$ and $Y$ we calculated the entropies $H(X)$ and $H(Y)$ according to the formulas in Section \ref{sec:hypotest}; in this case we use an alternative (equivalent) formulation for the mutual information $I(X;Y)$:
\begin{displaymath}
\nonumber I(X;Y) = \sum_{x \in X} \sum_{y \in Y} p(x,y) \log _2 \frac{p(x,y)}{p(x)p(y)}
\end{displaymath}

\begin{figure}
\centering
\includegraphics[trim=2.2cm 1.2cm 4.5cm 2cm,clip,width=\columnwidth]{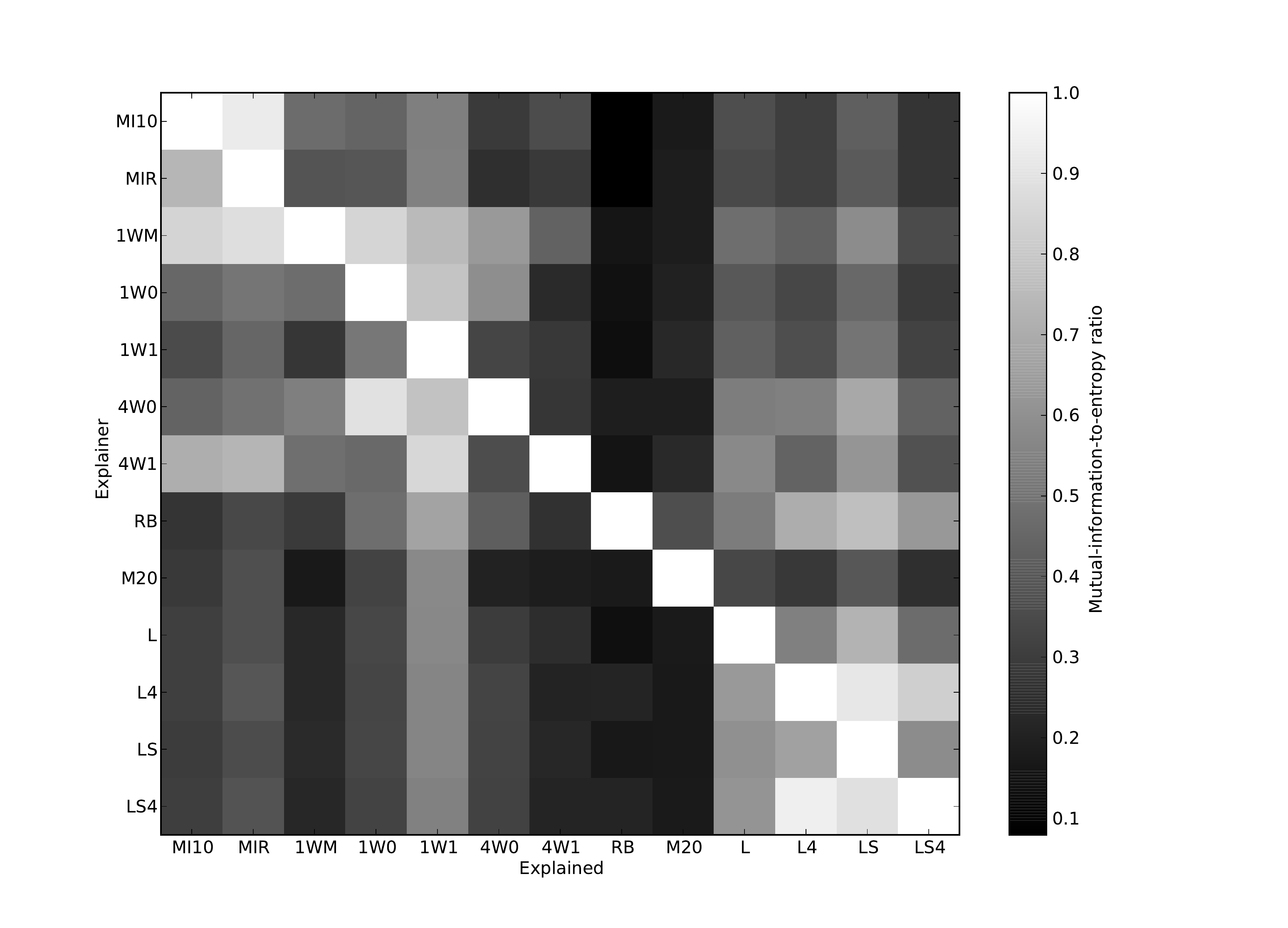}
\caption{Mutual information-to-entropy ratios for each test pair. Each entry is the fraction of the entropy of the test in that column explained by the test in that row. Brighter squares indicate that more of the variance of the explained test is explained by the explainer test. Test codes defined in Section \ref{sec:testMI}.}
\label{fig:testMI}
\end{figure}

The entropy $H(X)$ can be interpreted as the uncertainty in $X$, as measured by the number of bits required by an optimal code to specify a value from the distribution $p_X(x)$. The mutual information $I(X;Y)$ can be interpreted as the reduction in uncertainty in $X$ caused by knowledge of the value of $Y$, or vice versa (mutual information is symmetric) \cite{CoverThomas}. Figure \ref{fig:testMI} shows the ratio of $I(X;Y)$ to $H(X)$ for all tests $X$ and $Y$ used in MemtestG80; this ratio is the fraction of the uncertainty in $X$ explained by knowledge of $Y$. In Figure \ref{fig:testMI}, the $Y$ (the ``explaining'' distributions) are along the rows; the $X$ (the ``explained'' distributions) are along the columns. The following codes are used to refer to tests within MemtestG80:
\begin{description}
\item[MI10] Moving inversions, ones and zeros. Writes constant patterns of all-ones and all-zeros to memory.
\item[MIR] Moving inversions, random. Writes a constant (host-chosen) pseudorandom number to memory.
\item[1WM] Memtest86 variant of walking 1-byte test pattern.
\item[1W0/1] True walking zeros/ones pattern, 1-byte width.
\item[4W0/1] True walking zeros/ones pattern, 4-byte width.
\item[RB] Random blocks. Writes a different pseudorandom number (generated on the GPU) to each memory block.
\item[M20] Modulo-20 test. Described in Section \ref{sec:validation}.
\item[L, L4] Logic test, one or four iterations through LCG cycle.
\item[LS(4)] Logic test as L/L4, but with intermediate LCG state stored in shared memory rather than registers.
\end{description}

Several interesting trends emerge from this data:
\begin{enumerate}
\item \textbf{The Modulo-20 test stands on its own}\\
Both the M20 column and the M20 row have small values across their lengths, indicating the Modulo-20 test covaried strongly with no other test. This is likely due to the Modulo-20 test's increased sensitivity relative to other tests and reinforces the notion that it probes a different failure mechanism than do other tests.
\item \textbf{The Random Blocks test is a good logic test}\\
Although it was not intended as a logic test, the large values in the RB row for the columns corresponding to the LCG-based logic tests indicate that RB does a good job of capturing the errors measured by the LCG tests. Conversely, the small values in the RB column for the LCG tests demonstrate that RB is measuring a superset of errors relative to the LCG tests. This result is reasonable in retrospect: the RB test is very shader-logic intensive. We have designed it around a multithreaded, multi-core Park-Miller Minimal Standard pseudorandom number generator \cite{Park88}, which in the course of generating a new random number for each memory location performs many more logic operations than any other MemtestG80 test.
\item \textbf{The logic tests measure a distinct failure mode from most memory tests}\\
The four-iteration variants of the logic test (L4 and LS4) are poorly explained by most memory tests, and in particular, are less-well-explained by the memory tests than are their one-iteration counterparts (L and LS). This is to be expected, as the one-iteration variants are more influenced by memory errors. However, the bright block in the bottom-right of the mutual information plot shows that the logic tests covary strongly among themselves. Furthermore, memory tests have higher mutual information to the L4 test than the LS4 test, indicating that the use of shared memory in the logic test is a significant variable. Together, these results show that the logic tests detect a failure mode distinct from that tested in the memory tests, and that apparent logic errors can be triggered by soft errors in the on-GPU shared memory.
\end{enumerate}

\section{Conclusions}

We have presented the first large-scale study of error rates in GPGPU hardware, conducted over more than 20,000 GPUs on the Folding@home distributed computing network. Our control experiments on consumer-grade and dedicated-GPGPU hardware in a controlled environment found no errors. However, our large-scale experimental results show that approximately two-thirds of tested cards exhibited a pattern-sensitive susceptibility to soft errors in GPU memory or logic, confirming concerns about the reliability of the installed base of GPUs for GPGPU computation. We have further demonstrated that this nonzero error rate cannot be adequately explained by overclocking or time of day of execution (a proxy for ambient temperature). However, it appears to correlate strongly with GPU architecture, with boards based on the newer GT200 GPU having much lower error rates than those based on the older G80/G92 design. While we cannot rule out user error, misconfiguration on the part of Folding@home donors, or environmental effects as the cause behind nonzero error rates, our results strongly suggest that GPGPU is susceptible to soft errors under normal conditions on non-negligible timescales. 

Our negative control results suggest (but do not conclusively prove) that with environmental control and the use of dedicated-GPGPU hardware, GPGPU can be reliable. However, our experimental results raise concerns about the reliability of GPGPU on consumer-level hardware as installed in the wild. These data are particularly relevant both to GPU-based distributed computing applications and to vendors of consumer-targeted software that relies on GPU acceleration, such as recent video-encoding applications. We emphasize that although our data were collected only on NVIDIA GPUs, we have no reason to believe that the reliability picture would be significantly different for GPUs from ATI, Intel, Via, or other manufacturers, as the driving forces behind GPU development to date have not emphasized the strict reliability concerns found in GPGPU applications.

We have furthermore presented the design and validation of MemtestG80, our custom code to test NVIDIA CUDA-enabled GPUs for memory errors. We have released this tool under an open-source LGPL license at \texttt{https://simtk.org/home/memtest} in the hope that it can be used by others for GPU stress testing or self-checking.

\subsubsection*{Is dedicated GPGPU hardware the answer?}

Our work suggests several future avenues of investigation. One question which our sampling is unable to adequately answer is whether professional-level and GPGPU-dedicated boards are significantly more reliable than consumer-grade GPUs. While the underlying architectures are identical to those in consumer-grade cards, this specialist hardware is marketed as being more capable than consumer hardware, and NVIDIA has suggested that Tesla boards are recommended for mission-critical applications. 

While we were unable to sample a large enough number of Quadro and Tesla cards in our experiment to provide a conclusive answer to this question, the results of our negative control experiment (Section \ref{sec:validation}) suggest that the Tesla line may truly be more reliable than consumer hardware. In the course of our control experiment, we accumulated approximately 1.48 million MemtestG80 iterations over 8 Tesla boards. Each test operated over 5 times the typical amount of memory tested by a Folding@home MemtestG80 iteration, so the control experiment was equivalent to approximately 7.4 million Folding@home tester iterations, or over 925,000 per card. Neither any memory errors nor any logic errors were ever observed. Because the empirical probability of having a zero-error card in the Folding@home dataset is approximately one-third, the probability that our control Tesla cards were drawn independently from the same distribution is $\left( \frac{1}{3} \right)^8$, or less than 0.01\% (even less if drawn from the G80-only dataset). While our data do not rule out the possibility that environmental factors are at work (machine room versus uncontrolled environment) or that we tested an unusually good batch of cards, they suggest that the Tesla line are in fact more reliable than consumer-grade hardware.

\subsubsection*{What can be done?}

An obvious first step for software developers is to incorporate active memory test functionality, like that found in MemtestG80, to proactively detect malfunctioning cards. On the hardware side, the addition of parity or ECC functionality to the memory subsystem would guard against memory-induced silent errors. While the addition of partial ECC to the GDDR5 specification is a first step, it is not an end-to-end system, and cannot protect against errors in the memory controller or RAM itself. 

While memory testing and ECC can guard against errors in GPU memory, our logic tests indicate that transient errors on the GPU itself must also be considered. To combat these sources of faults, it may be necessary to implement measures such as redundant computation, advanced software error-detection \cite{Nicolescu03, Nicolescu01} or hardware redundancy \cite{Sheaffer07} mechanisms. 

Nevertheless, our data demonstrate that it is certainly possible to perform reliable GPGPU computing on consumer-grade hardware, but that doing so requires close attention to the characteristics of the hardware. With the great power of TFLOP-scale computation on a single board comes a great responsibility for the developer to ensure data integrity.

\section*{Acknowledgments}
We foremost thank all the donors on the Folding@home network, without whose donated computer time this study would not have been possible. We further thank Dr. Paul Coteus of IBM for discussions regarding soft error detection and correction mechanisms, Ewen Cheslack-Postava of Stanford for discussions on GPU architectures and the idea of an iterated logic tester, and Adam Beberg, Ewen Cheslack-Postava, Philip Guo, Peter Kasson, and Alex Rasmussen for helpful comments on the manuscript. ISH gratefully acknowledges support from an NSF graduate fellowship. We acknowledge support from NIH (R01-GM062868, U54 GM072970) and NSF (CHE-0535616).

\bibliographystyle{abbrv}
\bibliography{ihaque}
\end{document}